# The Bitcoin price formation: Beyond the fundamental sources

Jamal Bouoiyour[†] and Refk Selmi[‡]

**Abstract:** Much significant research has been done to investigate various facets of the link between Bitcoin price and its fundamental sources. This study goes beyond by looking into least to most influential factors -across the fundamental, macroeconomic, financial, speculative and technical determinants as well as the 2016 events- which drove the value of Bitcoin in times of economic and geopolitical chaos. We use a Bayesian quantile regression to inspect how the structure of dependence of Bitcoin price and its determinants varies across the entire conditional distribution of Bitcoin price movements. In doing so, three groups of determinants were derived. The use of Bitcoin in trade and the uncertainty surrounding China's deepening slowdown, Brexit and India's demonetization were found to be the most potential contributors of Bitcoin price when the market is improving. The intense anxiety over Donald Trump being the president of United States was shown to be a positive determinant pushing up the price of Bitcoin when the market is functioning around the normal mode. The velocity of bitcoins in circulation, the gold price, the Venezuelan currency demonetization and the hash rate were found to be the fundamentals influencing the Bitcoin price when the market is heading into decline.

[†] CATT, University of Pau, Avenue du Doyen Poplawski, 64016 Pau Cedex, France. E-mail : jamal.bouoiyour@univ-pau.fr
[‡] University of Tunis, Campus Universitaire, Avenue 7 Novembre, 2092, Tunis, Tunisia; University of Pau, France. Email: s.refk@yahoo.fr

# 1. Introduction

2016 was an eventful year. The two bastions of Anglo-Saxon capitalist democracy- the United Kingdom and the United States- have witnessed political earthquakes in the form of Brexit and the Trump's win in US presidential elections. Add to this the China economic slowdown that makes a bigger dent in the global economic outlook. Uncertainties will still be greater in 2017. In particular, the political issues will be omnipresent: important electoral events are scheduled to occur in Europe in 2017, including France, Germany and Netherlands, which may have a significant impact on the European political and economic lines for the coming years, without overlooking the Brexit consequences. Once Article 50 is triggered, and with the process of the United Kingdom's withdrawal from the European Union, the new European governments will develop their own negotiating strategies in the face of British demands. Because investing needs few unknowns, less uncertainty, visibility and trust, there is bountiful evidence that the increased fears over the results of presidential elections and doubt about the direction of future policies will make financial markets very volatile. This has led to a trend towards questioning the effectiveness of standard economic and financial structures which govern the conventional monetary and financial system. Here, the digital currency (in particular, Bitcoin) is leading the charge by providing a completely decentralized secure alternative to fiat currencies during times of economic and geopolitical upheaval. Bitcoin –which lives outside the confines of a single country's politics– currently profited from the increased uncertainty and the loss of faith in the stability of banking system and future economic security.

Although Bitcoin has been frequently discussed on various financial blogs and even mainstream financial media, the research community remains focused on the currency's safety and legal issues as well as the macroeconomic and financial aspects. However, the discussion about the Bitcoin response to the global uncertainties is still relatively sparse. Throughout this study, we tackle the price of the Bitcoin from a large perspective; we focus on the determinants of the price fluctuations in turbulent times, ranging from fundamental, speculative and technical sources to the 2016 events. While the determinants Bitcoin price have generated extensive debates over the last years (Buchholz et al. 2012; Kristoufek 2013; Bouoiyour and Selmi 2015; Ciaian et al. 2016, Bouri et al. 2017, etc.), most analyses ignore the fact that the impact of independent variables could fluctuate throughout the distribution of Bitcoin price. Although commonly applied regressions focus on the mean, deviations from the regression line can greatly affect the fit of the ordinary least squares (OLS). Median estimators and more general quantile estimators are generally less impacted by outlying observations in the response variable conditional on the covariates (Koenker and Bassett 1978; Konker 2005). Further, it is important to recognize that covariates can have an influence on the dispersion of the response variable as well as its location (i.e., heteroskedasticity).When this occurs, quantile regression unlike the OLS or the mean regression provides a more flexibility of covariate effects. While the traditional frequentists' approach to quantile regression has been largely used around asymptotic theories, not much research has been developed under the Bayesian framework (Kottas and Gelfand 2001). This paper seeks to address this matter by using Bayesian quantile regression (BQR). The use Bayesian quantile regression is based on at least four novelties. First, such regressions provide detailed and complete explanations of the determinants of Bitcoin price fluctuations, as Binder and Coad (2011) noted that the focus on mean effects could distort the relevant coefficient estimates, or might even fail to identify significant relationship. Second, according to Koenker and Hallock (2001), the estimator shows robustness to outliers on the dependent variable, making the quantile regression more efficient than OLS (ordinary least squares) regression in the context of non-normal error terms. Third, it is robust to skewness, heteroscedasticity and misspecification errors as it detects the underlying dependence structure between the examined time series, which could prove to be important as acknowledged that Bitcoin price displays successive ups and downs and thus nonlinear dynamics (Bouoiyour and Selmi 2016). Fourth, the BQR accounts for nonlinearity of the link function, the discontinuity of the loss function and the location and scale restrictions needed for parameter identification. Moreover, the use of quantile regression while controlling for endogeneity bias constitutes another contribution of this study.

The conducted methodology has allowed us to determine the least-to-most influential factors explaining Bitcoin price evolution. In particular, we distinguish three main groups of determinants. A large part of Bitcoin price buoyancy was attributed to the increased usefulness of Bitcoin as a transaction tool, the loss of trust in the Chinese yuan and the great uncertainty surrounding Brexit and India's demonetization at bull regime (upper quantiles). The uncertainty surrounding the Trump's victory in US elections was found to be the determinant surging the price Bitcoin's value at the normal mode (centrally located quantiles). Ultimately, the limited supply, the gold price, the announcement of demonetization in Venezuela and the hash rate were shown to be the driving forces of Bitcoin at bear state (bottom quantiles). These obtained findings highlight the importance of looking beyond the average correlation and the ability of the BQR method to capture the salient features in the correlation dynamics between Bitcoin price and its fundamentals.

The remainder of this article is organized as follows: Section 2 includes a brief discussion of Bitcoin determinants. Section 3 describes the data and the conducted methodology, while Section 4 discusses the results. Section 5 controls for endogeneity bias. Finally, Section 6 concludes.

## 2. The determinants of Bitcoin price

The past few years have witnessed considerable research concerning the Bitcoin price dynamics, and much has been written on the properties of this digital money. Unlike the fiat currencies (dollar, euro and yuan), Bitcoins are digital coins which are decentralized, not issued by any government or legal entity and not redeemable for gold or any other commodity. Bitcoins rely on cryptographic protocols and a distributed network of users to mint, store, and transfer. Instead, investors perform their business transactions themselves without any intermediary. According to the existing literature (Grinberg 2011; Buchholz et al. 2012; Kristoufek 2013; Bouoiyour and Selmi 2015, Balcilar et al. 2017, among others), Bitcoin price is determined by different factors (i) fundamental, macroeconomic and financial sources, (ii) speculation, and (iii) technical contributors. This study adds to these determinants the potential influence of the 2016 events.

According to Buchholz et al. (2012) and Ciaian et al. (2016), one of the key drivers of Bitcoin price is the interaction between supply and demand on the Bitcoin market. Although demand is primarily driven by its value as a medium of exchange, the supply is determined by the velocity of bitcoins in circulation, which is publicly known and predefined in the long-term. Kristoufek (2013) and Bouoiyour and Selmi (2015) pointed out the potential role of global macroeconomic and financial development -captured by variables such as exchange rates, exchange –trade ratio and gold prices- in determining Bitcoin price evolution. It must be stressed that the impact of macroeconomic and financial indicators on Bitcoin price may work through several channels. Among these channels, one can indicate that favorable macroeconomic and financial conditions may improve the use of Bitcoin in trade and exchanges and thus stimulate its demand which may exert positive influence on Bitcoin's value. Also, a fall in the prices of gold - normally perceived in theory as a hedge and safe haven to protect against several risks and to deal with ongoing volatility- may allow Bitcoin price to sustain its climb. If traders and investors lose trust in the yellow metal as a store of value, they may resort to Bitcoin. Recently, various studies argued the valuable role of Bitcoin as a hedge or safe haven (Eisl et al. 2015; Baur et al. 2015; Dyhrberg 2015; Popper 2015; Bouri et al. 2016). Bitcoin has been shown to be negatively correlated with stock prices, pointing toward its hedging capabilities. Dyhberg (2015), for example, tested the hedging capabilities of Bitcoin. The study documented that Bitcoin possess hedging characteristics as gold and can be incorporated in a portfolio to mitigate the harmful effects of sudden shocks. Another possible driver of the Bitcoin price is its speculative bahavior. Indeed, a rise in the attention toward this digital currency, accompanied with a way of actually investing in it, leads to an increase in the demand of Bitcoin and then to a surge of its prices. Accordingly, Lee (2014) showed that the alteration of positive and negative news contributed to high Bitcoin price cycles. This means that the attractiveness to Bitcoin via social networking can have a significant impact on Bitcoin price dynamics positively or negatively, depending to the type of news that dominate in the media at a specific time. In this ground, after the announcement of demonetization in India and the devaluation of the Venezuelan bolivar as a reaction to the demonetization government's decision on 11 December 2016, the attention to Bitcoin in India and Venezuela increased markedly, searching an alternative assets (in particular, Bitcoin). Moreover, the emergence of Bitcoin has provided new approaches concerning payments such as the "hash rate". It is an indicator of the processing power of the Bitcoin network. For security goal, the latter must make intensive mathematical operations that lead to an increase in the hash rate itself connected with an increase in cost demands for hardware. This would influence Bitcoin purchasers and then increase the demand of Bitcoin and in turn their prices (Bouoiyour and Selmi 2015). Furthermore, there are claims that events happening on 2016 have a significant impact on Bitcoin price. Some of the extreme price increases in the Bitcoin prices do coincide with dramatic events such as China economic downturn and the deterioration of Yuan[3], the Brexit vote on 23 June 2016 and the Trump's victory in US presidential elections on 08 November 2016 (see Figure 1).

---

[3] As most of the world's Bitcoin is mined and traded in China, it is expected that the current economic downturn will have a positive effect on Bitcoin's value. Analysts claimed that the latest Bitcoin's surge on the start of 2017 was due to choppy Chinese stock markets which were trading lower. The Chinese yuan has also fallen continuously against the US dollar in 2016 which has given a boost to Bitcoin price.

*Figure 1. The Bitcoin price fluctuations*

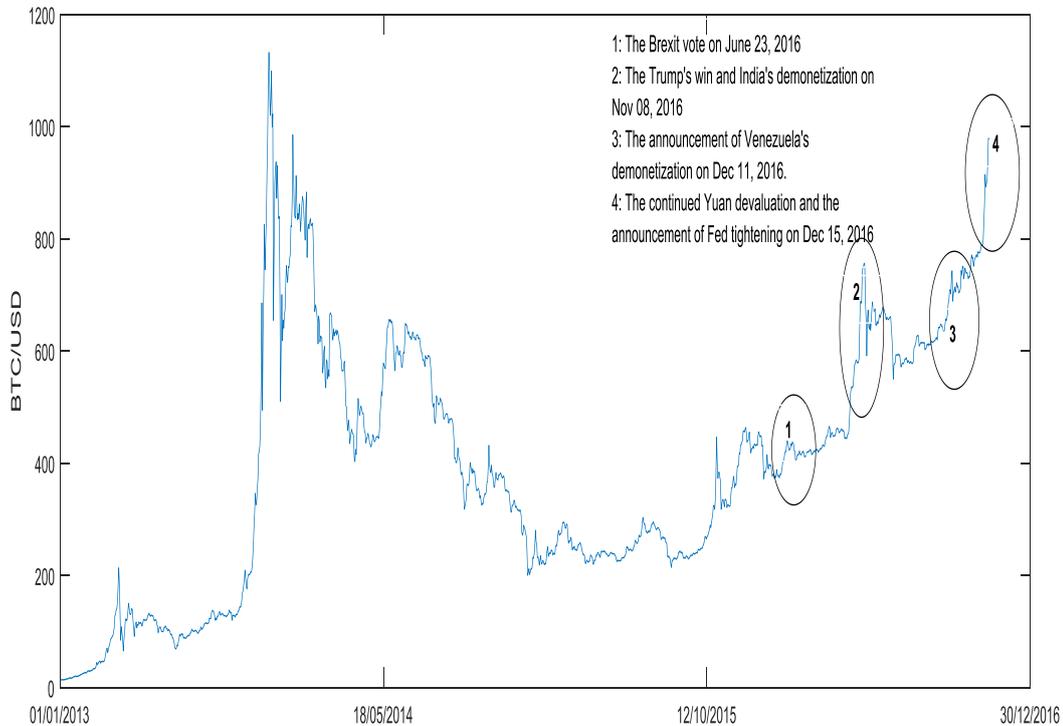

Source: CoinDesk.

## 3. Methodology and data

We, first, conduct a preliminary analysis aimed at plotting Bitcoin price against its determinants (Figure A, Appendix). We notice that the demand-supply fundamentals (the velocity of Bitcoins in circulation as proxy), the increased usefulness of Bitcoin as a transaction tool (exchange-trade ratio as indicator) and the loss of faith in fiat currencies (in particular, the deterioration of Chinese Yuan), play the lead role in explaining the recent increase in the price of Bitcoin. In fact, the increasing demand for a limited supply of Bitcoin raises cost per unit. As the Chinese currency enters a path of depreciation, investors will consider investing in digital currency that can preserve value and hedge risks. However, the recent Bitcoin price's climb does not appear highly interacted with the gold price fluctuations and the difficulty of processing power network. It is important to add here that the contribution of speculation appears also pronounced. During the last two weeks of December 2016, Google searches for the keyword "Bitcoin" in India and Venezuela rose by approximately 115, 250 and 400 percent, respectively. Venezuelans, for instance, have started to allocate their capital which is rapidly losing value mainly due to the deterioration of the Venezuelan bolivar in Bitcoin mining, to obtain a legitimate currency in Bitcoin to purchase goods internationally. Moreover, the 2016 global instability has directed funds to the Bitcoin market, but to a less extent. Specifically, the uncertainty surrounding the Brexit costs and the Donald Trump triumph have highlighted the Bitcoin glitter as investors look for a hedge or a safe haven.

As a side remark, we shall notice that that the distributions of some variables are broken, while others are less. This means that the average can vary greatly depending on the sample used. Standard linear regression techniques display the average dependence between a set of regressors and the dependent. This provides only a partial view of the focal linkage, as we might be interested in depicting the relation at different points in the conditional distribution ofdependant variable. The estimate of the mean is in this case compromised. It is therefore difficult to model it. Quantile regression (QR) provides this capability (D'Haultfoeuille and Givord 2014). Since the seminal work of Koenker and Bassett (1978), the number of researches on quantile regression raised remarkably. Applications of quantile regression emerge in many research areas, ranging from ecology over genetics to economics and finance. QR continues to be an interesting tool as it accounts for a set of regression curves that differ across distinct quantiles of the conditional distribution of the dependent variable, and then overcomes the various problems that OLS is confronted with. In general, error terms are not constant across a distribution, thereby violating the axiom of homoscedasticity. In fact, the error is assumed to have accurately the same distribution whatever values may be taken by the components of the vector of dependant variable. The latter affects only the location of the conditional distribution of dependant variable, but not its scale. Quantile regression has emerged as a good supplement to ordinary mean regression. As the mean gives an incomplete picture of a single distribution, the regression gives also incomplete picture for a set of distributions. It seems valuable and useful, since it provides a more accurate description of changes than focusing solely on the mean. The upper or lower quantiles of the response variable may depend on the covariates very differently from the center. For better comprehensive analysis on quartile regressions, you can refer to Koenker (2005). In addition, by concentrating on the mean as a measure of location, informations about the tails of a distribution are lost. Although OLS can be inefficient if the errors are heavily non-normal, QR is more robust to non-normal errors and outliers (Koenker and Bassett 1978; Koenker and Xiao 2002). QR also offers a richer characterization of the data, enabling us to take into account the impact of a covariate on the entire distribution of the dependent variable, not merely its conditional mean. Furthermore, OLS is sensitive to extreme outliers, which can distort the findings.

In brief, a QR is suited to determine how evolve time series for all portions of a probability distribution (i.e., slopes from the minimum to the maximum responses). However, some specific features such as skewness, fat-tails, outliers, breaks, truncated-censored data, and heteroskedasticity, can sometimes shadow the nature of the dependence among the variables under study and the covariates, so that the conditional mean would not be the most appropriate statistic to effectively understand the nature of the investigated interdependence. One of the most challenging issues in quantile regression analyses is related to the fact that this model involves minimal assumptions (i.e., the error distribution) that may lead to non-normal errors. A satisfactory inference procedure is difficult to be tackled, since the asymptotic covariance matrix of quantile estimates normally make us an unknown error density function, which cannot be estimated reliably. In a Bayesian quantile regression framework, we can thus efficiently deal with this problem. There exist several studies on quantile regression both in frequentist and Bayesian framework, dealing with parametric and non-parametric approaches. For a detailed review, one can refer for instance to Lum and Gelfand (2012). In this paper, we will conduct a Bayesian approach to quantile regression.

This technique possesses considerable advantages compared to the usual quantile regression estimates. First, Markov Chain Monte Carlo (MCMC) method can be easily carried out to obtain the posterior distributions even in "complex" situations. Second, Bayesian quantile regression performs appropriately when the conditional distribution is not symmetric. Even though quantile regression problem may be resolved by minimizing the objective function, the Bayesian approach to quantile regression must specify precisely likelihood. The asymmetric Laplace (ASL) distribution that prompts equivalence between posterior mode and simple quantile regression estimates has been carried out to construct Bayesian quantile regression model (Yu and Moyeed 2001). Given this, a specific distributional assumption for the error terms has been defined (Yue and Rue 2011):

$$y_i = x_i^{'}\beta_\tau + \varepsilon_{i\tau}, i = 1,...,n \qquad (1)$$

For the $\tau^{th}$ conditional quantile function, $0 < \tau < 1$, the $\hat{\beta}(\tau)$ is called the $\tau^{th}$ regression quantile and defined as a solution to the following problem:

$$\hat{\beta}(\tau) = \arg\min_{\beta \in \Re} \sum_{i=1}^{n} \rho_\tau(y_i - x_i\beta) \qquad (2)$$

where $\rho_\tau$ is a weight calculated by $\rho_\tau(y) = y(\tau - I(y<0))$, I (.) is the indicator function and $0 < \tau < 1$.

Following the ASL distribution, errors are independent and identically distributed, i.e., ($\varepsilon_{i\tau} | \sigma^2$) i.i.d ASL ($0, \sigma^2, \tau$), with the following density:

$$p(\varepsilon_{i\tau} | \sigma^2) = \frac{\tau(1-\tau)^2}{\sigma} \exp\left(-\omega_\tau(\varepsilon_{i\tau}, 0) \frac{|\varepsilon_{i\tau}|^2}{\sigma}\right) \tag{3}$$

Then, the error distributions yield $y_i | \beta_\tau, \sigma^2 \sim \text{ASL}(x_i' \beta_\tau, \tau)$ such that the density of the responses is denoted as:

$$p(y_i | \beta_\tau, \sigma^2) = \frac{\tau(1-\tau)^2}{\sigma} \exp\left(-\omega_\tau(y_i, x'\beta_\tau) \frac{|y_i - x_i'\beta_\tau|^2}{\sigma}\right) \tag{4}$$

While the ASL distribution enables to properly express quantile regression within Bayesian framework, it may lead to more complicated inference based on Markov Chain Monte Carlo simulation. To avoid such complexity, the ASL can be represented as a scale mixture of normal distributions, as following:

$$(y_i | z_i, \beta_\tau, \sigma^2) \sim N(x_i' \beta_\tau + \varepsilon z_i, \sigma^2 / \omega_i) \tag{5}$$

where $\varepsilon = \frac{1-2\tau}{\tau(1-\tau)}, \omega_i = \frac{1}{\delta^2}, \delta^2 = \frac{2}{1-\tau}$

The Bayesian inference can be applied effectively by imputing the scale variables $z_i$ as supplementary unknowns (Fahrmeir et al. 2013). Then, the evolution of estimator may be observed by setting τ = 0.5 (median regression). The first and the last quantiles are obtained by setting τ = 0.1 and τ = 0.9, respectively.

As mentioned above, the present research attempts to identify the least-to-most influential Bitcoin fundamentals across the fundamental, macroeconomic and financial determinants (the velocity of bitcoins in circulation: VC; the exchange – trade ratio: ETR; the gold price: GP), the speculative factors (the increased interest in Bitcoin in Venezuela and India), the technical drivers (the hash rate), and the events occurring in 2016 (the deterioration of Chinese currency: Yuan, the British and U.S. VIX: BV and USV, respectively). Here, we provide a detailed description of all the analyzed variables together with their source links.

The model to be estimated is given by:

$$B\hat{P}I_t^\tau = \hat{\omega}(\tau) + \hat{\delta}_1(\tau)VC_t + \hat{\delta}_2(\tau)ETR_t + \hat{\delta}_3(\tau)GP_t + \hat{\delta}_4(\tau)ABI_t + \hat{\delta}_5(\tau)ABV_t + \hat{\delta}_6(\tau)HR_t$$

$$\hat{\delta}_7(\tau)YUAN_t + \hat{\delta}_8(\tau)BV_t + \hat{\delta}_9(\tau)USV_t \quad (6)$$

where $B\hat{P}I_t^\tau$ is the estimated $\tau$-conditional quantile of Bitcoin price, and the estimated parameters $\hat{\delta}_k(\tau)$ for k=1, …, 9 are functions of $\tau$.

The Bitcoin price index (BPI) is an index of the exchange rate between the US dollar (USD) and the Bitcoin (BTC). The CoinDesk Bitcoin Price Index represents an average of Bitcoin prices across leading bitcoin exchanges. The total number of bitcoins in circulation is given by a known algorithm until it reaches 21 million bitcoins. The velocity of bitcoins in circulation (VC) is, by definition, the frequency at which one unit of each currency is used to purchase tradable or non-tradable products for a given period. As a measure of the transactions use, we employ the ratio between trade and exchange transactions volume or the ratio between volumes on the currency exchange markets and in trade, which we abbreviate to trade-exchange ratio (ETR). Although it is not easy to distinguish between several incentives of internet users searching for information about the keyword "Bitcoin", Google searches can be a valuable tool to predict the Bitcoin market (Kristoufek 2013); millions of users daily interact with search engines, creating valuable sources of data regarding various aspects of the world. While the frequency of searches of a specific keyword is incomparable to a sentiment index, it can provide partial information which can be used to understand a complex phenomenon. Besides, the creation of new bitcoins is mainly determined by difficulty that mirrors the computational power of Bitcoin miners (hash rate: HR). As proxies of The uncertainty surrounding the Brexit and the Trump's victory, we use the British and US implied or realized volatility indices (BV and USV[4], respectively) that have the advantage of being directly observable, and thus appear more objective as measures of uncertainty over such event. It must be stressed at this stage that the volatility index is a sentiment indicator that allows determining when there is too much optimism or pessimism in the market. Also, we should point out that BV and USV respond sensitively to all events (reflecting both economic and geopolitical issues) that may cause uncertainty, and the China's economic slow-moving, the Brexit, the U.S. presidential election results, the announcement of demonetization in India and Venezuela and the plunge of gold price are no exception. Due to data availability, we analyze the relationships starting from 01 January 2015 to 30 December 2016, which in turn, gives us a total of 729 observations. Table 1 reports all the data used and their sources.

---

[4] As the data of VIX for India and Venezuela are unavailable for the same period of study, we have chosen to use another uncertainty proxy also largely used in the behavioral finance literature which is the Google Trends.

*Table 1. Data sources*

|  | **Variables** | **Definition** | **Sources** |
|---|---|---|---|
| The dependent variable | BPI | Bitcoin price index | CoinDesk (www.coindesk.com/price) |
| Fundmantal, marcroeconomic and financial determinants | VC | Velocity of Bitcoin | Blockchain (http://www.blockchain.info) |
|  | ETR | Exchange Trade Ratio | Blockchain(http://www.blockchain.info) |
|  | GP | Gold price | quandl website |
| Speculation | ABI | Attention to Bitcoin in India | Google Trends (http://trends.google.com) |
|  | ABV | Attention to Bitcoin in Venezuela | Google Trends (http://trends.google.com) |
| Technical drives | HR | Hash rate | Blockchain (http://www.blockchain.in |
| 2016 events | Yuan | The Chinese Yuan | DataStream of Thomson Reuters |
|  | BV | British VIX | DataStream |
|  | USV | U.S. VIX | DataStream |

## 4. Discussion of results

An initial step consists of using OLS regression to have initial information about the determinants of Bitcoin price evolution. The idea here is to have a case of benchmarking to compare the OLS with BQR approach. The OLS results reported in Table 2 indicate that the majority of coefficients of the independent variables are insignificant. Only the coefficients of VC, ABI and BV seem significant. There exist also sharp differences between the conditional median (i.e., LAD) and the mean (i.e., OLS) estimates. While ABV and USV exert a positive and significant influence on Bitcoin price when $\tau=0.5$ (LAD), insignificant effects were found when accounting for OLS findings. Likewise, although the VC's coefficient appears insignificant when $\tau=0.5$, it seems positive and significant when considering OLS estimates. This can be attributed to the fact that the mean effect of the independent variables on the dependent variable may be under or over estimate impacts or even fail to properly determine full possible influences (Cade and Noon 2003); hence the need to perform more elaborate methods, in particular Bayesian quantile regression that brings accurate information about the average dependence between variables and the upper and the lower tails. Table 2 displays the BQR estimates for the relationship between Bitcoin price and its fundamentals. We look at the Bitcoin from various aspects that might affect its prices ranging from the fundamental, the macroeconomic, the financial, the speculative and the technical contributors to the influence of the global uncertainties surrounding the 2016 events.

### 4.1. Fundamental, macroeconomic and financial determinants

The results reported in Table 2 reveal that the money supply –proxied by the velocity of bitcoins (VC) in circulation- exerts a negative impact on Bitcoin price in bear state. Specifically, the VC coefficient fluctuates between -0.14 ($\tau=0.2$) and -0.15 ($\tau=0.4$). This result is consistent with the quantity theory putting in evidence that the price of Bitcoin decreases with the stock of bitcoins. The money supply works as a standard supply so that its increase leads to price decrease. We should mention at this stage that Bitcoin faces a great challenge with respect its limited amount recording 21 million units in 2140, implying that the money supply would not increase after this date. In addition, we note that the exchange-trade ratio is positively and strongly correlated with the price of Bitcoin when the market is bullish (upper quantiles). In particular, the ETR coefficient varies among 0.30 ($\tau=0.6$) and 0.39 ($\tau=0.9$). The usage of Bitcoin in real transactions (purchases, services, etc.) is significantly connected to the fundamental aspects of its value. Theoretically, the price of the currency should be positively related to its usage for transactions, as it raises the utility of holding the currency leading to an increase in its prices. Moreover, a negative and modest correlation between gold price and Bitcoin price was found in bear state (bottom quantiles; the GP coefficient ranges between -0.004 when $\tau=0.1$ and -0.001 when $\tau=0.2$). Bitcoin and gold do not evolve in the same direction. As the two assets are viewed as hedge and safe haven in turbulent time, we can mention that one causes the other; but the factors driving the price of Bitcoin and the price of gold may be dissimilar.

### 4.2. Speculation

Using Google search queries for two countries (India and Venezuela), we document that the growing attention in Bitcoin leads to increasing prices when the market is functioning around the normal and the bull regimes. The increased interest toward Bitcoin in India and Venezuela contributes positively and significantly to Bitcoin price. Precisely, an increase by 10% in the attention towards Bitcoin in India raises the price of Bitcoin by 0.2% when τ=0.4 and by approximately 1.4% when τ=0.8. By delving into the case of Venezuela, we note that a climb of the interest in Bitcoin by 10% increases BPI by about 1.3% when τ=0.3 and by about 1.5% when τ=0.5. The government of India shocked its citizens on 08 November 2016 when announcing the demonetization of the 500 and 1000 rupee notes. Because more than 90% of transactions use physical currency in India, the influence of demonetization may be undoubtedly life-changing for the entire Indian population. Since the announcement of demonetization, Indians want to park their black money (in old currency notes) in Bitcoins. Also, the economy of Venezuela has been shred apart by the financial collapse in response to the government's decision to demonetize the nation's biggest denomination banknote prompting a deterioration of the bolivar exchange rate. This has pushed Venezuelans to buy Bitcoin with hopes to obtain an alternative currency, driving up the Bitcoin price.

### 4.3. Technical drivers

Our results indicate a negative effect of the hash rate on Bitcoin price at bear (bottom quantiles; τ=0.1, τ=0.2 and τ=0.3) and normal states (τ=0.4 and τ=0.5); such influence decreases by moving from bear to normal regime. Particularly, an increase by 10% in the hash rate increases the price of Bitcoin by about 1% when τ=0.1, while it surges it by about 0.6% when τ=0.5. The more miners that join the Bitcoin network, the greater the network hash rate is. Mining can be perceived as a kind of investment towards Bitcoin (Ciaian et al. 2016). A strong hash rate connected with growing cost demands for hardware and electricity push miners to the mining pool. If these miners employ the coins as an alternative to the direct investment, they can turn to Bitcoin purchasers and thus amplify the demand for Bitcoin and thus its prices (Kristoufek 2013).

### 4.4. The 2016 events

Our findings document that the 2016 events play a potential role in explaining the Bitcoin price variation. In particular, we note that the deterioration of yuan against dollar is negatively and strongly correlated to the price of Bitcoin at bullish regime (upper quantiles; τ=0.7, τ=0.8 and τ=0.9). In 2016, China saw its foreign-exchange reserves collapse by approximately 8 percent. This sharp decrease of reserves has arisen as China's currency plunged by about 6 percent against the dollar. During times of market turbulence, it is acknowledged that there is a tendency towards "flight to safety". The negative correlation between Yuan and BPI implies that Bitcoin has acted as a hedge against the depreciation of yuan[5]. In general, Bitcoin is a hedge two scenarios: tightened capital controls and the market anxiety. Currently, investors in China see both of those happening: The People's Bank of China (PBOC) cracked down with stricter capital controls, and the price of the yuan has collapsed markedly against the dollar during the 12 months of 2016. Moreover, Bitcoin price responds positively and significantly to the volatility witnessing Britain in the onset of Brexit and USA after the announcement of Trump's victory in the presidential elections. This holds at upper quantiles (bullish states). An increase by 10% in BV leads to a rise by 0.9% when $\tau=0.6$ and by about 1.2% when $\tau=0.8$. Expectedly, a significant number of Britain and US residents are not happy with the outcome of the vote and the elections, respectively, which has created a potent nervousness in the market, in turn, diminishing the confidence among traders and investors. Soon after the Brexit results, market participants started to trade in the US dollar, which they believe won't be damaged as much as the euro or pound in the onset of Brexit. But Bitcoin –which lives outside the confines of a single country's politics – has also gained a sharp validity with Brexit news. Regarding the 2016 U.S. elections, a rise by about 10% in USV surges the Bitcoin price by about 1% when $\tau=0.3$, by 1.2% when $\tau=0.5$ and by approximately 0.8% when $\tau=0.8$. The financial markets had widely priced in a win for Clinton, who they viewed as a better short-run outcome because she represented few unknowns and thus less uncertainty. Donald Trump's victory has sent U.S. markets on a tumultuous ride. Markets are reacting as investors find out how heavy are the president-elect's statements on trade, fiscal policy and regulation. The uncertainty the Trump's agenda can push people to hoard alternative assets such as Bitcoin. While it remains unclear what to expect, there appears to be a quite general consensus in the Bitcoin community that whatever Trump's policies turn out to be, Bitcoin will benefit[6].

---

[5] A strong (weak) safe haven is defined as an asset that has a significant positive (negative) return in periods where another asset is in distress, while hedge has to be negatively correlated (uncorrelated).

[6] The reason that is making Bitcoiners hopeful about Trump's win is the inclusion of Bitcoin supporters like Peter Thiel, Balaji Srinivasan and Mick Mulvaney in his team. Peter Thiel is a technology entrepreneur and investor; he is the co-founder of PayPal, a Bitcoin enthusiast and has invested into multiple Bitcoin companies. Balaji Srinivasan is one of the best-funded Bitcoin startups so far. He is the co-founder and CEO of 21; the latter has developed a full stack set of technologies for practical Bitcoin micropayments. Also, Mick Mulvaney, the designated Director of the Office of Management and Budget under Trump's presidency, is viewed as one of the most representatives of the crypto community since he is more outspoken about Blockchain technology and Bitcoin. Having Bitcoin believers in the Trump's team is a win for the Bitcoin community in the whole.

Furthermore, we use the Koenker and Xiao (2002) test to evaluate whether the estimated quantile regression relationships are conform to the location shift hypothesis which assumes the same slope parameters for all of the conditional quantile functions. The Koenker and Xiao (2002) test computes that all the covariate effects satisfy the null hypothesis of equality of the slope coefficients across quantiles. In particular, the difference between slope estimates at the τ and (1- τ) quantiles is examined. The rejection of the null hypothesis implies that the magnitude of the slope coefficient, estimated at the different parts of the return distribution is different and that the difference is significant. Our results indicate that the rejection favors the Bayesian quantile regression for the link between Bitcoin price and all its fundamentals.

*Table 2. The Bayesian quantile regression results*

|  | Quantile | Coefficient | Prob. |
|---|---|---|---|
| *OLS Results* | | | |
| OLS(VC) |  | 0.04983* | 0.0426 |
| OLS(ETR) |  | 0.00813 | 0.1549 |
| OLS(GP) |  | -0.15649 | 0.1174 |
| OLS(ABI) |  | 0.06782** | 0.0024 |
| OLS(ABV) |  | 0.02649 | 0.2589 |
| OLS(HR) |  | 0.14962 | 0.1038 |
| OLS(Yuan) |  | 0.06462 | 0.1058 |
| OLS(BV) |  | 0.12085* | 0.0735 |
| OLS(USV) |  | 0.10567 | 0.1389 |
| *Fundamental macroeconomic and financial determinants* | | | |
| VC | 0.100 | -0.014368 | 0.2087 |
|  | 0.200 | -0.147722* | 0.0577 |
|  | 0.300 | -0.15362* | 0.0518 |
|  | 0.400 | -0.153742* | 0.0607 |
|  | 0.500 | -0.010411 | 0.3869 |
|  | 0.600 | -0.009174 | 0.4134 |
|  | 0.700 | -0.006099 | 0.5664 |
|  | 0.800 | -0.013953 | 0.1364 |
|  | 0.900 | -0.012633 | 0.1733 |
| ETR | 0.100 | -0.002368 | 0.9961 |
|  | 0.200 | 0.022154 | 0.9320 |
|  | 0.300 | 0.217671 | 0.2961 |
|  | 0.400 | 0.173979 | 0.1376 |
|  | 0.500 | 0.170676 | 0.1512 |
|  | 0.600 | 0.302331* | 0.0208 |
|  | 0.700 | 0.369179** | 0.0027 |
|  | 0.800 | 0.367999*** | 0.0007 |
|  | 0.900 | 0.393178*** | 0.0001 |
| GP | 0.100 | -0.004155* | 0.0524 |
|  | 0.200 | -0.001242** | 0.0079 |

|  | 0.300 | 1.03E-05 | 0.9995 |
|---|---|---|---|
|  | 0.400 | 0.004357 | 0.8272 |
|  | 0.500 | -0.002200 | 0.9318 |
|  | 0.600 | -0.003016 | 0.9142 |
|  | 0.700 | -0.003802 | 0.2804 |
|  | 0.800 | 0.000177 | 0.9945 |
|  | 0.900 | 0.025053 | 0.2455 |
| *Speculation* | | | |
| ABI | 0.100 | 0.003977 | 0.7388 |
|  | 0.200 | 0.006705 | 0.7319 |
|  | 0.300 | 0.013517 | 0.1384 |
|  | 0.400 | 0.020439* | 0.0243 |
|  | 0.500 | 0.082919** | 0.0010 |
|  | 0.600 | 0.138696*** | 0.0008 |
|  | 0.700 | 0.127745* | 0.0336 |
|  | 0.800 | 0.148288* | 0.0516 |
|  | 0.900 | 0.115335* | 0.0570 |
| ABV | 0.100 | 0.014156 | 0.3817 |
|  | 0.200 | 0.032073 | 0.1068 |
|  | 0.300 | 0.131183* | 0.0336 |
|  | 0.400 | 0.133455* | 0.0674 |
|  | 0.500 | 0.152838* | 0.0287 |
|  | 0.600 | 0.06284 | 0.1088 |
|  | 0.700 | 0.04513 | 0.1187 |
|  | 0.800 | 0.025473 | 0.1599 |
|  | 0.900 | 0.020209 | 0.1913 |
| *Technical drivers* | | | |
| HR | 0.100 | 0.107624** | 0.0023 |
|  | 0.200 | 0.094851* | 0.0331 |
|  | 0.300 | 0.083228*** | 0.0001 |
|  | 0.400 | 0.080806*** | 0.0003 |
|  | 0.500 | 0.068808*** | 0.0004 |
|  | 0.600 | 0.065919 | 0.1181 |
|  | 0.700 | 0.033436 | 0.4030 |
|  | 0.800 | 0.017052 | 0.6914 |
|  | 0.900 | 0.031473 | 0.5438 |
| *The 2016 events* | | | |
| Yuan | 0.100 | -0.136140 | 0.4883 |
|  | 0.200 | -0.136607 | 0.1298 |
|  | 0.300 | -0.118562* | 0.1980 |
|  | 0.400 | -0.117092 | 0.1273 |
|  | 0.500 | -0.108550 | 0.4664 |
|  | 0.600 | -0.118404 | 0.1677 |
|  | 0.700 | -0.195816* | 0.0870 |
|  | 0.800 | -0.137067** | 0.0051 |
|  | 0.900 | -0.142893*** | 0.0004 |
| BV | 0.100 | 0.103726 | 0.2345 |
|  | 0.200 | 0.127470 | 0.1867 |

|  | 0.300 | 0.127002 | 0.2358 |
|---|---|---|---|
|  | 0.400 | 0.122871** | 0.0011 |
|  | 0.500 | 0.134276*** | 0.0009 |
|  | 0.600 | 0.096330*** | 0.0000 |
|  | 0.700 | 0.120694*** | 0.0000 |
|  | 0.800 | 0.124113*** | 0.0000 |
|  | 0.900 | 0.117350*** | 0.0000 |
| USV | 0.100 | 0.001033 | 0.9559 |
|  | 0.200 | 0.039846 | 0.2386 |
|  | 0.300 | 0.099151*** | 0.0000 |
|  | 0.400 | 0.100448*** | 0.0000 |
|  | 0.500 | 0.121399*** | 0.0000 |
|  | 0.600 | 0.146810*** | 0.0000 |
|  | 0.700 | 0.086599*** | 0.0005 |
|  | 0.800 | 0.084870* | 0.0216 |
|  | 0.900 | 0.010673 | 0.3109 |
| C | 0.100 | 3.851660 | 0.2251 |
|  | 0.200 | 3.716549* | 0.0502 |
|  | 0.300 | 2.097040* | 0.0322 |
|  | 0.400 | 2.348100* | 0.0103 |
|  | 0.500 | 2.261169* | 0.0167 |
|  | 0.600 | 2.084521* | 0.0405 |
|  | 0.700 | 1.962397* | 0.0387 |
|  | 0.800 | 2.047476* | 0.0169 |
|  | 0.900 | 1.498318* | 0.0723 |
| Koenker and Xiao (2002) test | | | |
| OLS(VC) and BQR | 0.0013** | | |
| OLS(ETR) and BQR | 0.0010** | | |
| OLS(GP) and BQR | 0.0008*** | | |
| OLS(ABI) and BQR | 0.0003*** | | |
| OLS(ABV) and BQR | 0.0012** | | |
| OLS(HR) and BQR | 0.0009*** | | |
| OLS(Yuan) and BQR | 0.0004*** | | |
| OLS(BV) and BQR | 0.0008*** | | |

Notes: ***, ** and * imply significance at the 1%, 5% and 10% levels, respectively.

By comparing the BQR results with the simple regression findings reported in Figure A (Appendix), we clearly note that the coefficients of some independent variables (like gold price and Yuan) are quite far from the BQR (in particular, the median estimate) and also the OLS results summarized in Table 2. An element that may explain these heterogeneous results is that we plot the dependent variable (Bitcoin price) in function of one-by-one explanatory variables (unconditional data analysis), while with BQR and OLS we regress BPI on several determinants (conditional data analysis). Studying the bivariate relationship may not be robust when some relevant independent variables are not included. When we consider only two variables, we generally fall on the problem of simple regression without control variable which is unable to capture proper results with regard to the nexus studied since it may distort the estimate. Further, the median regression (LAD) is more robust to outliers than OLS regression and simple linear regression in general, as it avoids assumptions about the parametric distribution of the error process (Baum 2013).

## 5. The issue of endogeneity

In perfect markets, the Bitcoin price falls with the velocity and the stock of Bitcoins, but raises with the exchange trade ratio (transactions) and the size of Bitcoin economy (Bitcoins in circulation). Note that in the market equilibrium, the Bitcoin price, the exchange trade ratio, and the total stock of Bitcoins in circulation adjust simultaneously, which may generate endogeneity issues when estimating the relationship between Bitcoin price and these fundamentals (Ciaiain et al. 2016). In standard regression model (for example OLS), the endogeneity of simultaneous variables may violate the exogeneity assumption of a regression equation. The estimation of asymmetric interdependencies among interdependent time series in the presence of mutually correlated variables is subject to endogeneity problem (Lütkepohl and Krätzig 2004). It should be remembered that the use of the BQR does not solve the problem of endogeneity that remains. To control for possible endogeneity bias, there are many methods including GMM, two-stage least squares (2SLS) method and instrumental variable (IV) regression. We decided here to apply 2SLS for at least two main reasons: First, GMM requires differentiability of the moment functions, while Bayesian quantile regression consists on non-differentiable sample moments. This implies that the combination of these two methods can be inappropriate. Second, for instrumental quantile regression, it turns out very difficult to find proper instruments with regard to the relationship between Bitcoin price and the exchange trade ratio or the total stocks of Bitcoins in circulation. To this end, we use all the independent and the lagged dependent variables to calculate the estimated values of the exchange –trade ratio variable. Then, these estimated values are used in place of the actual values of ETR.

We begin this analysis by carrying out a simple 2SLS regression to compare its results with those of 2SLS within BQR. The findings summarized in Table 3 reveal that there are some differences between BQR-based 2SLS (the conditional median) and simple 2SLS estimates. For example, we note that ABI and HR have an insignificant coefficient when applying 2SLS, and positive and significant coefficients when using BQR-based 2SLS. Also, the yuan is likely to have a negative impact on BPI when considering 2SLS results, while it have no effect when using BQR-based 2SLS.This dissimilarity can be partially due to the asymmetry of the conditional density and to the strong effect exerted on the least squares fit by the possible outlier observations in the sample.

By controlling for endogeneity, we note modest changes in the groups of Bitcoin determinants (Table 3). In particular, the uncertainty surrounding the Trump's victory (USV) joins the variables strongly explaining the Bitcoin's value when the market is at the bullish regime (first group, Table 2), and the intense fears over the announcement of demonetization in Venezuela (ABV) joins the second group accounting for the variables pushing up the Bitcoin price when the market is at the normal state.

Using the Koenker and Xiao (2002), we note that the null hypothesis of equal slope is rejected at the conventional significance levels for all the cases, indicating that the slope coefficient of the different Bitcoin determinants differs at the various parts of the return distribution.

*Table 3. The Bayesian quantile regression results after controlling for endogeneity bias*

|  | Quantile | Coefficient | Prob. |
|---|---|---|---|
| **2SLS Results** | | | |
| 2SLS(VC) |  | -0.10496 | 0.1078 |
| 2SLS(ETR) |  | 0.20672 | 0.3945 |
| 2SLS(GP) |  | -0.00243* | 0.0595 |
| 2SLS(ABI) |  | 0.06974 | 0.1148 |
| 2SLS(ABV) |  | 0.07892* | 0.0836 |
| 2SLS(HR) |  | 0.09425 | 0.2367 |
| 2SLS(Yuan) |  | -0.11842** | 0.0093 |
| 2SLS(BV) |  | 0.2586 | 0.5411 |
| 2SLS(USV) |  | 0.1439 | 0.8765 |
| *Fundamental, macroeconomic and financial determinants* | | | |
| VC | 0.100 | -0.161937** | 0.0011 |
|  | 0.200 | -0.129492*** | 0.0009 |
|  | 0.300 | -0.134944*** | 0.0006 |
|  | 0.400 | -0.140194** | 0.0013 |
|  | 0.500 | -0.118532 | 0.1064 |
|  | 0.600 | -0.118536 | 0.1578 |
|  | 0.700 | -0.115900 | 0.1683 |
|  | 0.800 | -0.111354 | 0.1674 |
|  | 0.900 | -0.177685 | 0.1595 |
| ETR | 0.100 | 0.490723 | 0.1696 |
|  | 0.200 | 0.338552 | 0.1626 |

|  | 0.300 | 0.456242 | 0.9232 |
|  | 0.400 | -0.315345 | 0.9792 |
|  | 0.500 | 0.289688 | 0.6562 |
|  | 0.600 | -0.024949 | 0.7197 |
|  | 0.700 | 0.401490 | 0.3006 |
|  | 0.800 | 0.377211** | 0.0064 |
|  | 0.900 | 0.414345* | 0.0158 |
| GP | 0.100 | -0.007395* | 0.0682 |
|  | 0.200 | 0.082668 | 0.8870 |
|  | 0.300 | -0.005958 | 0.9906 |
|  | 0.400 | -0.005325 | 0.2243 |
|  | 0.500 | -0.001916* | 0.0614 |
|  | 0.600 | -0.583889 | 0.2492 |
|  | 0.700 | -0.000285** | 0.0011 |
|  | 0.800 | -0.654508 | 0.1544 |
|  | 0.900 | -0.987736 | 0.1202 |

*Speculation*

| ABI | 0.100 | 0.058608*** | 0.0002 |
|  | 0.200 | 0.013217*** | 0.0000 |
|  | 0.300 | 0.108062 | 0.1178 |
|  | 0.400 | 0.020718*** | 0.0004 |
|  | 0.500 | 0.044954*** | 0.0001 |
|  | 0.600 | 0.082477*** | 0.0005 |
|  | 0.700 | 0.065346** | 0.0013 |
|  | 0.800 | 0.112214*** | 0.0009 |
|  | 0.900 | 0.117603** | 0.0010 |
| ABV | 0.100 | 0.080659 | 0.8564 |
|  | 0.200 | -0.284721 | 0.9913 |
|  | 0.300 | 0.155379** | 0.0056 |
|  | 0.400 | 0.169802** | 0.0048 |
|  | 0.500 | 0.105933 | 0.5032 |
|  | 0.600 | -0.234585 | 0.4713 |
|  | 0.700 | 0.213110 | 0.1985 |
|  | 0.800 | 0.556073 | 0.2356 |
|  | 0.900 | 0.923938 | 0.1475 |

*Technical drivers*

| HR | 0.100 | 0.123709* | 0.0571 |
|  | 0.200 | 0.105017** | 0.0022 |
|  | 0.300 | 0.066098 | 0.2521 |
|  | 0.400 | 0.095723* | 0.0249 |
|  | 0.500 | 0.068729** | 0.0064 |
|  | 0.600 | 0.080520 | 0.1772 |
|  | 0.700 | 0.238222 | 0.1112 |
|  | 0.800 | 0.288278 | 0.1421 |

*The 2016 events*

| Yuan | 0.100 | -0.167209 | 0.1567 |
|  | 0.200 | -0.122168 | 0.1549 |
|  | 0.300 | -0.122519 | 0.1963 |
|    | 0.400 | -0.146701    | 0.2341 |
|    | 0.500 | -0.167072    | 0.2576 |
|    | 0.600 | -0.173128*   | 0.0145 |
|    | 0.700 | -0.171860*   | 0.0206 |
|    | 0.800 | -0.183776*   | 0.0183 |
|    | 0.900 | -0.192882*   | 0.0100 |
| BV | 0.100 | 0.248498     | 0.8951 |
|    | 0.200 | 1.004760     | 0.9670 |
|    | 0.300 | 1.088995     | 0.9968 |
|    | 0.400 | 0.529520     | 0.8984 |
|    | 0.500 | 0.346673     | 0.9968 |
|    | 0.600 | 0.105373**   | 0.0023 |
|    | 0.700 | 0.130680     | 0.7864 |
|    | 0.800 | 0.104290**   | 0.0096 |
|    | 0.900 | 0.155169*    | 0.0289 |
| USV| 0.100 | 0.211360     | 0.9914 |
|    | 0.200 | 0.208108     | 0.9276 |
|    | 0.300 | 0.194220     | 0.9751 |
|    | 0.400 | 0.181202     | 0.9533 |
|    | 0.500 | 0.183290     | 0.9974 |
|    | 0.600 | 0.188873     | 0.1007 |
|    | 0.700 | 0.101502***  | 0.0004 |
|    | 0.800 | 0.188931     | 0.1167 |
|    | 0.900 | 0.116032**   | 0.0083 |
| C  | 0.100 | 5.146855     | 0.3067 |
|    | 0.200 | 4.983597     | 0.8878 |
|    | 0.300 | 5.986939     | 0.8917 |
|    | 0.400 | 6.986641     | 0.7699 |
|    | 0.500 | 7.219691     | 0.1773 |
|    | 0.600 | 8.108894*    | 0.0132 |
|    | 0.700 | 8.695493*    | 0.0238 |
|    | 0.800 | 8.150009**   | 0.0081 |
|    | 0.900 | 14.51538**   | 0.0023 |
| ST |       | 0.0325*      |        |
| SY |       | 0.1567       |        |

| Koenker and Xiao (2002) test | |
|---|---|
| 2SLS(VC) and BQR   | 0.0161*  |
| 2SLS(ETR) and BQR  | 0.0098** |
| 2SLS(GP) and BQR   | 0.0073** |
| 2SLS(ABI) and BQR  | 0.0113*  |
| 2SLS(ABV) and BQR  | 0.0046** |
| 2SLS(HR) and BQR   | 0.0105*  |
| 2SLS(Yuan) and BQR | 0.0087** |
| 2SLS(BV) and BQR   | 0.0033** |

Notes: ***, ** and * imply significance at the 1%, 5% and 10% levels, respectively; ST: Sargan-Hansen test; SY: Stock–Yogo weak identification test.

## 6. Concluding remarks

Since its creation in 2009, particular attention has been given to Bitcoin. Despite its popularity and gradual worldwide acceptance, most people are still confused as to what a Bitcoin actually is. The status of Bitcoin as an alternative currency, transactions tool or a speculative bubble is still subject to on-going debate. We were therefore compelled to revisit the issue of the determinants of Bitcoin price from larger perspective. Here, we contribute to the existing literature by searching the potential contributors of Bitcoin prices ranging from fundamental, macroeconomic, financial, speculative and technical sources to the most marked events of 2016. For this purpose, we apply a Bayesian quantile regression model while controlling for endogeneity bias. This technique is used for inference about the dynamic interdependencies between quantiles of the response distribution and available covariates.

The BQR results stretch out some relationships, which could have been difficult to detect using standard econometric methods (OLS, LAD or 2SLS). In particular, three main groups of Bitcoin determinants were found. The first group contains the most influential Bitcoin drivers when the market is at its bull state (the use of Bitcoin in trade and the yuan deterioration and the uncertainty surrounding the Brexit and India's demonetization). The second group is formed by the Bitcoin fundamentals when the market is at its normal mode (the uncertainty surrounding the 2016 U.S. presidential elections).The third group accounts for Bitcoin fundamentals when the market is at its bear state (the velocity of bitcoins, the gold price, the increased fears over the Venezuelan demonetization and the hash rate). Potentially, the Bitcoin fundamentals were also ranged from the least influential (the difficulty level of Bitcoin mining and the gold price) to the most influential contributors (the fundamental and speculative sources and the 2016 events).

Currently, the legal status of Bitcoin in many nations across the globe becomes known[7] (Figure B, Appendix), and many large companies are accepting bitcoins as a legitimate source of funds[8] . As demand increases and supply shortens, the Bitcoin price rises. We also deduce that Bitcoin and the dynamics of gold are likely to be moderately interdependent in bearish regimes; such dependence is expected as both assets are considered as safe haven in times of chaos.

---

[7] Figure B indicates that the rate of Bitcoin adoption becomes greater in 2016. While several nations are permitted to make payments in Bitcoin, there exist also other countries around the world are still keeping a close eye on this digital currency.

[8] 8912 businesses are referenced on coinmap on April 2017; more details are available in the following link: https://coinmap.org/#/world/50.80593473/-50.53710938/3

Furthermore, the Bitcoin prices increase with the growing interest to this cryptocurrency (i.e. speculation). Due to the much-publicized and vexed demonetization policy enforced by Indian and Venezuelan governments and the restricted movement of capital outside, Bitcoin has presented an attractive option to get a hold of cash. We further document that the Bitcoin price's surge motivates users to become miners. Interestingly, there are beliefs that events happening on 2016 are mainly behind the recent Bitcoin's bullish run. Our findings confirm this expectation by showing that the fears over the continued deterioration of yuan against U.S. dollar have pushed Chinese traders and investors to place their bets and investments in Bitcoin. In addition, the anxieties over the Brexit and 2016 U.S. presidential elections outcomes have encouraged investors to seek a secure alternative, contributing positively to the price of Bitcoin.

Although the Bitcoin's climb in response to the 2016 global uncertainties underscores a confidence in Bitcoin as a safe haven, hedge and an alternative currency, experts are still reluctant to give this volatile virtual currency such status. Investors and traders are generally interested in hedges that mitigate the volatility of their portfolio, but also they are likely interested in buying some sort of insurance against extreme tail events. Bitcoin has several properties that make it a very interesting asset in both cases. Currently, the loss of faith in the stability of banking system and the future economic security worsened, and market uncertainty heightened across the globe. However, Bitcoin which lives outside the confines of a single country's politics has profited from the recent ongoing volatility. These properties may justify that Bitcoin serves as a hedge in turbulent times. But from a legal perspective, Bitcoin does not appear to share the characteristics of traditional safe-haven investments[9]. Even though Bitcoin is a liquid asset even in times of market upheaval, it is a high-risk, volatile and speculative investment. The Bitcoin market is also too narrow and not mature enough to be integrated into global financial markets.

Last but not least, we demonstrate the importance to look beyond the average correlation, and to keep up with the market' behavior (bullish, normal or bearish states) when analyzing the determinants of Bitcoin price.

---

[9] When market turmoil arises, investors are known to sell "risky" assets and buy "safer" assets, also known as "flights to safety" (Baele et al. 2015)

# Appendix

*Figure A.  Simple regression of Bitcoin price on its fundamentals*

*Fundamental, macroeconomic and financial determinants*

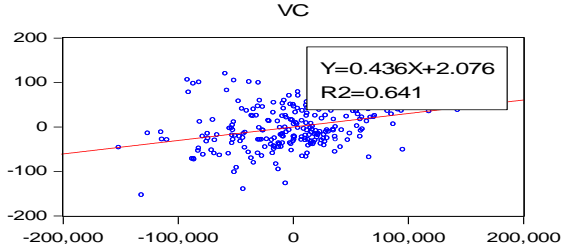

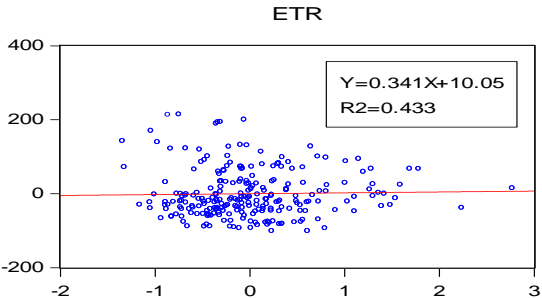

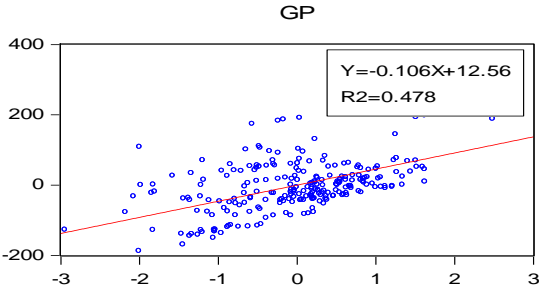

*Speculation*

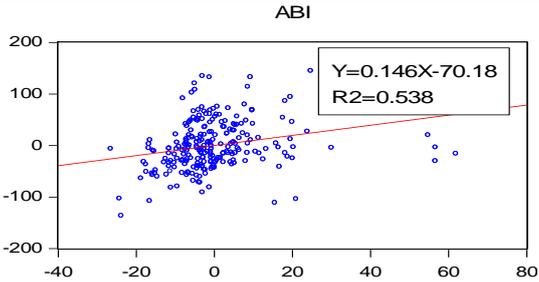

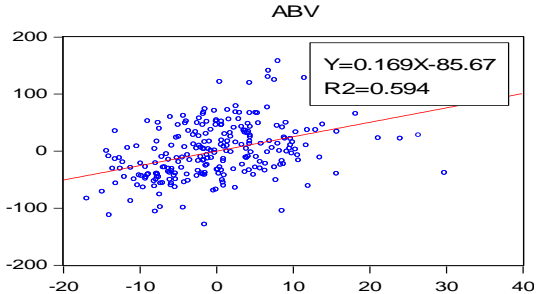

## Technical drivers

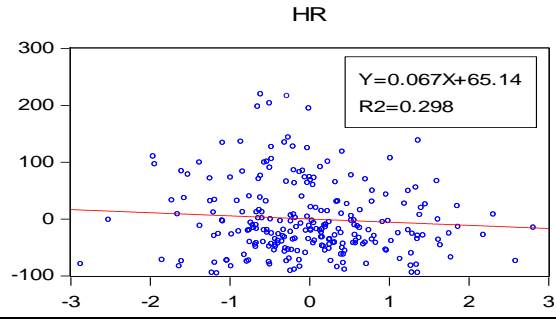

## The 2016 events

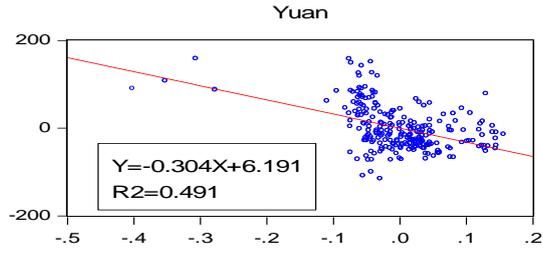

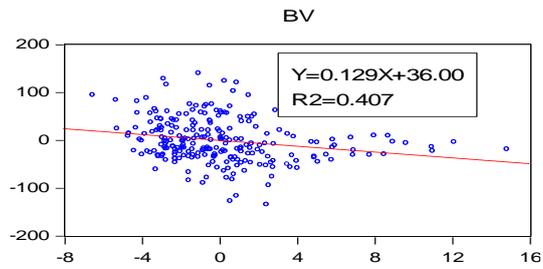

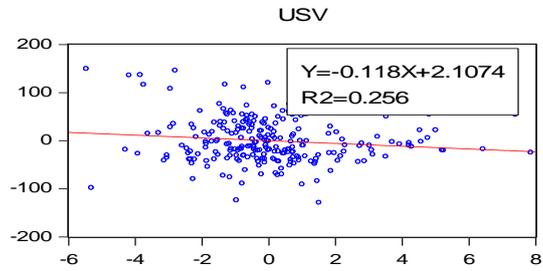

*Figure B. The Bitcoin adoption by country in 2016*

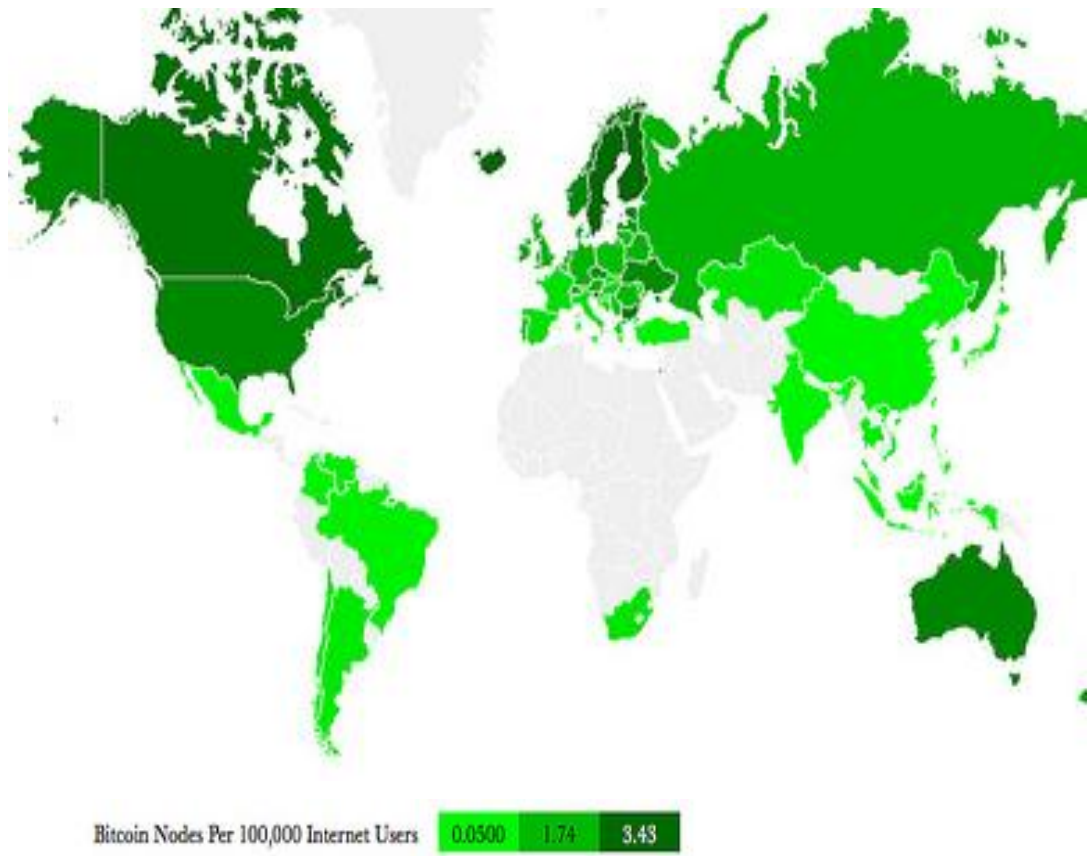

Source: bitcointalk.org